\documentclass[11pt,letterpaper,english,aps]{revtex4}
\usepackage{times}
\usepackage[T1]{fontenc}
\usepackage{amsmath}
\usepackage{graphicx}

\makeatletter
\usepackage{babel}
\makeatother
\begin{document}

\title{In silico design of metal-dielectric nanocomposites for solar energy applications }

\author{Justin Trice$^{a,b}$, Hernando Garcia$^{c}$, Radhakrishna Sureshkumar$^{b,d}$, Ramki Kalyanaraman$^{a,b}$}

\affiliation{$^{a}$Department of Physics, Washington University, St. Louis, Missouri 63130, USA;}

\affiliation{$^{b}$Center for Materials Innovation, Washington University, St. Louis, Missouri 63130, USA;}

\affiliation{$^{c}$Department of Physics, Southern Illinois University, Edwardsville, Illinois 62026, USA;}

\affiliation{$^{d}$Department of Energy, Environmental and Chemical Engineering, Washington University,
St. Louis, Missouri 63130, USA}

\begin{abstract}
Recently, a homogenization procedure has been proposed, based on the tight lower bounds of the
Bergman-Milton formulation, and successfully applied to dilute ternary nanocomposites to predict
optical data without using any fitting parameters {[}Garcia et al. Phys. Rev. B, 75, 045439 (2007)].
The procedure has been extended and applied to predict the  absorption coefficient of a quaternary
nanocomposite consisting of Cu, Ag, and Au nanospheres embedded in a SiO$_{\text{2}}$ host matrix.
Significant enhancement of the absorption coefficient is observed over the spectral range 350-800
nm. The magnitude of this enhancement can be controlled by varying the nanosphere diameter and
the individual metal volume fraction with respect to the host matrix. We have determined the
optimal composition resulting in enhanced broadband (350nm-800nm) absorption of the solar spectrum
using a simulated annealing algorithm. Fabricating such composite materials with a desired optical
absorption  has potential applications in solar energy harvesting. 
\end{abstract}
\maketitle

\section{INTRODUCTION}

Judicious selection of the materials used for a given photonic application is of paramount importance
in tailoring the desired optical properties of a composite material \cite{GarciaAPL06}. Light
waves, when directed to the interface between a metal and a dielectric are capable of a resonant
interaction with the mobile electrons present at the surface of the metal, giving rise to surface
plasmons, i.e. density waves of electrons that propagate along the interface \cite{Atwater07}.
Similarly, such nanocomposites offer much potential in designing materials with enhanced light
absorption due to the local electromagnetic field enhancement from the surface plasmon resonances
that occur at the metal-dielectric interface \cite{Sihvola99}. The wavelength-dependent absorption
of such metal-dielectric nanocomposites can be tuned by varying parameters such as the shape,
size and volume fraction of the nanoparticles as well as the dielectric constants of the host
and metal \cite{ZhangAPL07,DouillardJAP07,StuartAPL98}. The increasing emphasis on renewable
energy sources has generated interest in finding cost-effective ways of enhancing the efficiency
of solar energy harvesting systems. One promising route is the use of metal nanoparticles or
nanocomposite coatings as an absorbing layer on silicon devices, thereby giving rise to photocurrent
enhancement that correlates well with the plasmonic response of the applied coating \cite{SchaadtAPL05,PillaiJAP07,StuartAPL98}.
While consideration of the effects of electromagnetic scattering due to the coating morphology
on the device performance needs to be addressed \cite{PillaiJAP07}, optimization of the absorption
of incident solar energy though  design of an absorption coating is a critical step in the development
of improved solar absorbing materials.  Recently, we have developed and validated against experimental
data a hierarchical homogenization procedure capable of accurately predicting the absorption
characteristics of multi-metal/dielectric nanocomposites \cite{GarciaPRB07}. Typically, optical
analysis and experimentation in the literature are limited to nanocomposites with one or two
differing species \cite{BattaglinNucInsMethPhysB00,DouillardJAP07,MagruderNonCrystSold94,SchaadtAPL05,PillaiJAP07}.
Here, we show that by employing the above homogenization procedure each metal added in the dielectric
host introduces an added degree of freedom from which the optical response of a multi-metal nanocomposite
material may be tailored. In particular, we demonstrate how a Cu:Ag:Au-SiO$_{\text{2}}$ nanocomposite
might be used as an absorption coating optimized to the solar spectrum via the manipulation of
parameters such as volume fraction and particle size.  The ability to design such materials from
first principles can guide future experimental efforts aimed at controlling particle size and
spacing in techniques such as laser-induced dewetting \cite{Favazza05b,Favazza06b,TricePRB07},
ion implantation \cite{BattaglinNucInsMethPhysB00,MagruderNonCrystSold94}, and laser-assisted
chemical vapor decomposition \cite{ElihnJAP07,ElihnAppPhysA01}.

\section{HOMOGENIZATION PROCEDURE AND NANOCOMPOSITE SPECTRAL RESPONSE}

\begin{figure}[!tbh]
\begin{centering}\includegraphics[width=3in]{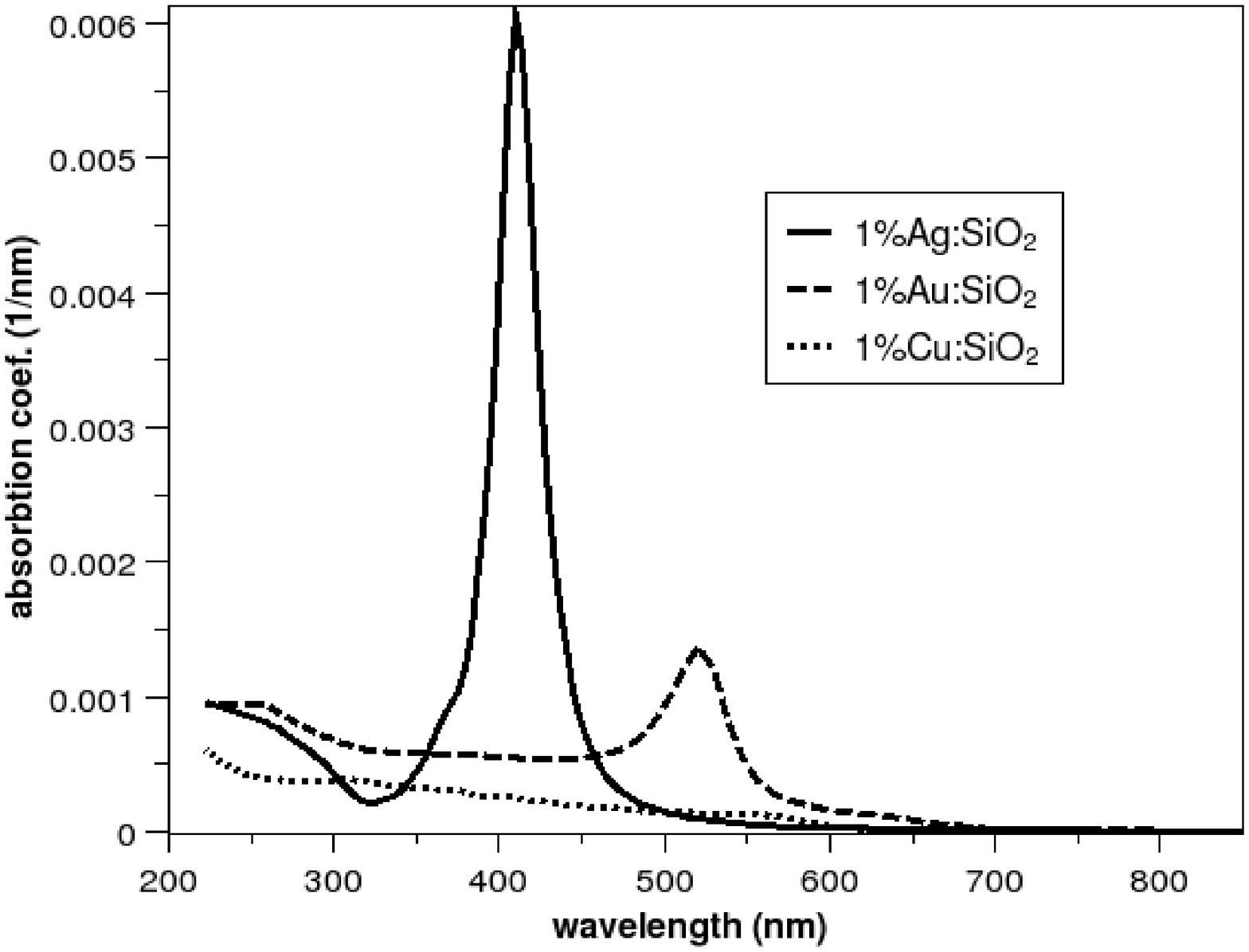}\par\end{centering}

\caption{\label{fig:singleres} The wavelength-dependent response of single species of 30 nm diameter
nanoparticles embedded in SiO$_{\text{2}}$. Ag and Au exhibit strong optical absorption at 414
nm and 529 nm respectively. Cu in SiO$_{\text{2}}$ yields a fairly broad resonance spanning
from 200 nm 645 nm. }
\end{figure}

The Bergman-Milton formula \cite{BermanPRL80,MiltonJAP81} for the effective dielectric constant
for a binary metal-dielectric composite is given by:

\begin{equation}
\epsilon_{eff}(\gamma)=\left[\frac{f_{a}}{\epsilon_{a}}+\frac{f_{h}}{\epsilon_{h}}-\frac{2f_{a}f_{h}\left(\epsilon_{a}-\epsilon_{h}\right)^{2}}{3\epsilon_{a}\epsilon_{h}\left[\epsilon_{h}\gamma+\epsilon_{a}\left(1-\gamma\right)\right]}\right]^{-1}\label{eq:BergMilt}\end{equation}
where $f$ corresponds to the component's volume fraction, $\epsilon$ is the component's dielectric
constant with subscripts $a$ and $h$ referencing the metal and host matrix respectively. $\gamma$
is a geometrical factor taking into account the shape of the metal particles. For the purposes
of this paper we are interested in spherical particles and accordingly choose $\gamma=\frac{2}{3}(1-f_{a})$.
Equation \ref{eq:BergMilt} denotes the effective dielectric function for the two component case
of a metal particle embedded in a host dielectric matrix. Note it is customary to let $f_{h}=1-\sum f_{n}$
where $f_{n}$ corresponds to the metal components. In Fig. \ref{fig:singleres} the effective
absorption coefficient, $\alpha=\frac{2\pi}{Re(\sqrt{\epsilon_{eff}})\lambda}Im(\epsilon_{eff})$
where $\lambda$ is the wavelength of incident electromagnetic energy, of a single metal species
of either Ag, Au, or Cu embedded in a dielectric SiO$_{\text{2}}$ matrix is presented.  The
effective absorption coefficients of Ag and Au in SiO$_{2}$ have the desirable property of exhibiting
strong absorption near their plasmonic resonances; although this response is localized at 414
nm and 529 nm for Ag and Au nanoparticles respectively. Cu embedded in SiO$_{\text{2}}$ exhibits
a rather broad spectral response spanning from 200 nm to 645 nm. However, the maximum obtainable
magnitude of the effective absorption of the Cu-SiO$_{\text{2}}$ system is clearly weaker than
that of Ag-SiO$_{\text{2}}$ and Au-SiO$_{\text{2}}$. Quantitatively, over the range of wavelengths
spanning from 225 to 1200nm, the Cu, Ag, and Au in SiO$_{\text{2}}$ systems in Fig. \ref{fig:singleres}
have an integrated absorption $I=\int_{225nm}^{1200nm}\alpha d\lambda$ of 0.229, 0.338, and
0.203 (unitless) respectively. Thus, Ag exhibits the most enhancement over the spectral range,
but this enhancement is localized and may not be the optimum response for a given application.

\begin{figure}[!tbh]
\begin{centering}\includegraphics[width=3in,keepaspectratio]{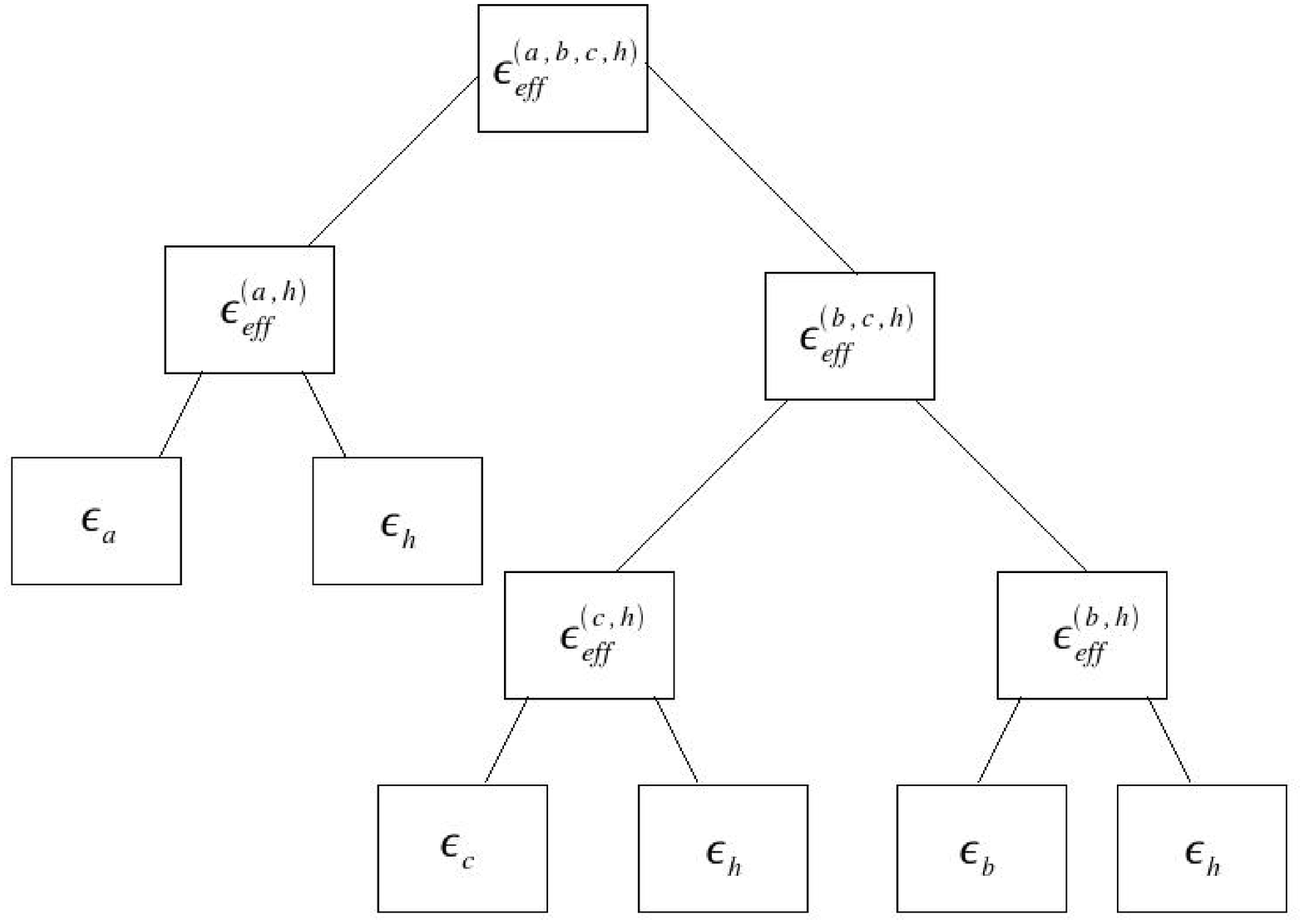}\par\end{centering}

\caption{\label{fig:mixschematic} Schematic of nanocomposite mixing rule to find effective dielectric
constant for the case of dilute quaternary metal-dielectric system.}
\end{figure}

For our proposed design, we limit the analysis to a dilute system (i.e. total volume fraction
of metal versus dielectric $\leq10\%$) because it is desirable from a practical standpoint of
cost and fabrication efficiency. A second constraint we have placed is to set the minimum nanosphere
diameter to be no less than $10\, nm$. Above this size regime, quantum confinement effects may
be neglected \cite{HalperinRevModPhys1986}. Moreover, nanoparticles of size $>$ 10 nm can be
assembled on SiO$_{2}$ substrates using robust nanomanufacturing techniques such as laser-induced
dewetting \cite{Favazza05b,Favazza06b,TricePRB07}, ion implantation \cite{BattaglinNucInsMethPhysB00,MagruderNonCrystSold94},
and laser-assisted chemical vapor decomposition \cite{ElihnJAP07,ElihnAppPhysA01}. In Fig. \ref{fig:mixschematic},
the homogenization procedure is presented for a dilute 3-metal mixture in a host dielectric matrix.
The top level represents the effective dielectric function for the target nanocomposite of design
interest. At each level, the total volume fraction is constrained so $f_{h}+\sum f_{n}=1$. The
schematic for homogenization of a 2-metal mixture and alloys in a host dielectric matrix has
been presented elsewhere \cite{GarciaPRB07}. Here, the 2-metal mixing rule is extended to treat
the quaternary system by viewing each component as having an effective permittivity $\epsilon_{eff}^{a,h}$,
$\epsilon_{eff}^{b,h}$, and $\epsilon_{eff}^{c,h}$. During the mixing process, the average
electric field within the composite is held constant while the final effective permittivity is
calculated using equal volumes at each level of mixing. As described in Ref. \cite{GarciaPRB07}:
the effective permittivity of an N-component mixture can be determined by mixing N-1 binary mixtures,
each comprising of a host and a distinct metal,  the host being common to the N-1 pairs. 

\begin{figure}[!tbh]
\begin{centering}\includegraphics[width=3in]{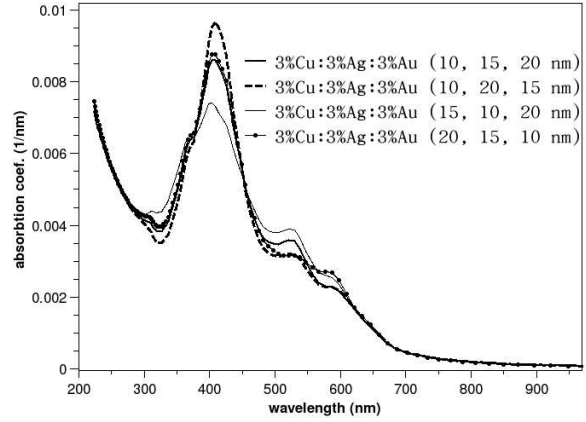}\par\end{centering}

\caption{\label{fig:constvfrac} Dependence of composite absorption on particle size calculated from
homogenization procedure. The volume fractions of the metal species are held fixed at 3\%. The
size of the nanoparticle diameters are permuted between 10 nm, 15 nm, and 20 nm. Decreasing the
diameter of a given metal broadens the spectral behavior of the metal's plasmon resonance at
the expense of absorption magnitude. }
\end{figure}

The nanoscale effect of change in the electron relaxation time due to scattering of electrons
on the metal-dielectric interface can be quantified via a modified Drude model \cite{YangApplPhysA96}
expressed by Eq. \ref{eq:teff}: 

\begin{equation}
\frac{1}{\tau_{eff}}=\frac{1}{\tau_{bulk}}+\frac{\nu_{F}}{2d_{n}}\label{eq:teff}\end{equation}
where $\tau_{eff}$ is the effective electron relaxation time, $\tau_{bulk}$ is the bulk electron
relaxation time, $d_{n}$ corresponds to the particle's diameter and $\nu_{F}$ is the Fermi
velocity of the electrons. The size effect may then be accounted for by modifying the imaginary
portion of the effective dielectric coefficient \cite{GarciaPRB07} as follows

\begin{equation}
Im(\epsilon_{a})=\frac{\omega_{p}^{3}}{\omega^{2}\tau_{eff}}=Im(\epsilon_{a}^{bulk})\left(\frac{2d_{n}+\nu_{F}\tau_{bulk}}{2d_{n}}\right)\label{eq:ImagCorr}\end{equation}
where $\omega_{p}$ is the plasmon frequency  and  $\omega$ is the angular frequency of the
incident electromagnetic energy. By exploiting this nanosize effect, one is able to broaden the
plasmonic peaks at the expense of their magnitude as demonstrated in Fig. \ref{fig:constvfrac}. 

\begin{figure}[!tbh]
\begin{centering}\includegraphics[width=3in,keepaspectratio]{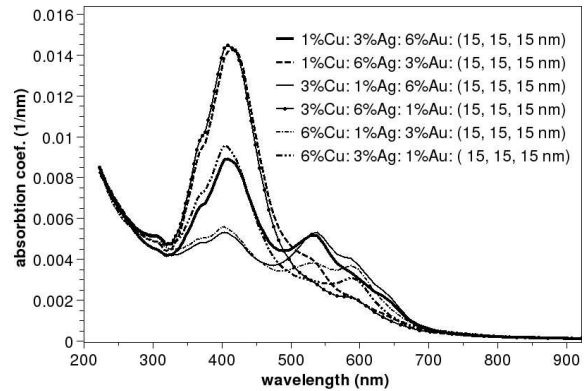}\par\end{centering}

\caption{\label{fig:constantpart} Dependence of absorption on volume fraction calculated from homogenization
procedure. The nanoparticles for all metal species are held fixed at 15 nm. The volume fraction
of each metal species is permuted between 6\%, 3\%, and 1\%. The ensuing nonlinear response of
the effective medium's absorption coefficient is shown.}
\end{figure}

\begin{figure}[!tbh]
\begin{centering}\includegraphics[width=3in]{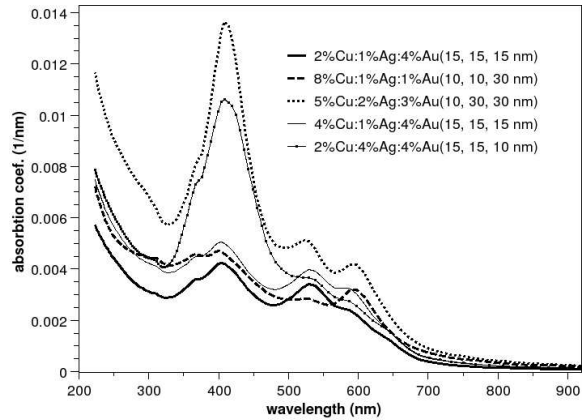}\par\end{centering}

\caption{\label{fig:control} Manipulation of broad-band absorption for  a dilute Cu:Ag:Au-SiO$_{\text{2}}$
mixture.  The only constraint is that the nanoparticle diameter lie between 10 nm and 30 nm.
The spectral response can be tailored to a given application fairly precisely over the spectral
range of 350-800nm.}
\end{figure}

Ag and Au embedded in SiO$_{\text{2}}$ exhibit very intense absorptions near their plasmonic
resonance. On the other hand, the spectral response of Cu is broadband,  extending into the  range
of green light. The net effect of mixing these particular materials allows one to achieve a wide
degree of freedom in choosing the  spectral response of the composite.  Fig.\ref{fig:constantpart}
shows the nonlinear response of the composite system as a function of   variations in the volume
fraction. Here the volume fraction of each metal is permuted between 6\%, 3\%, and 1\% while
the size of the particles is fixed at 15 nm. Changing volume fraction results in a nonlinear
change in the wavelength depended absorption. Further, decreasing the diameter of a given metal
broadens the spectral behavior of the metal's plasmon resonance at the expense of absorption
magnitude. Fig. \ref{fig:constvfrac} shows the trade-off between spectral broadening and plasmon
resonance magnitude. To demonstrate the wide degree of control of this particular quaternary
system, Fig. \ref{fig:control} presents examples of manipulating magnitude over the spectral
range as a designer might do for a given application. Our mixing approach predicts the intrinsic
spectral behavior of the nanocomposite with certain volume fraction ratios and particle diameters.
Extrinsic parameters, such as the actual composite volume and area exposed to solar irradiation,
determine the magnitude of solar energy absorbed.

\section{OPTIMIZATION}

\begin{figure}[!tbh]
\begin{centering}\includegraphics[width=3in]{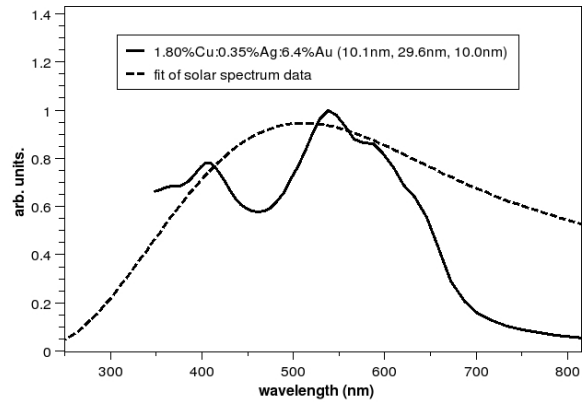}\par\end{centering}

\caption{\label{fig:optimizedspec} Optimized absorption of the solar spectrum for a Cu:Ag:Au-SiO$_{2}$
nanocomposite. The optimal composition of the system was found to be 1.80\%Cu:0.35\%Ag:6.40\%Au
with respective particle diameters of 10.1 nm, 29.6 nm and 10.0 nm. Spectral data curve fit is
plotted in arbitrary units for visualization. }
\end{figure}

Towards designing the initial framework of an solar-harvesting device, we have used a simulated
annealing algorithm \cite{KirkpatrickScience1983,MetropolisJChemPhys1953,PressNumRecC1997} to
determine the optimum composite configuration.   We define the {}``energy'' to be minimized
as $E(\mathbf{f},\mathbf{d})=\sum abs\left|(s_{i}-\alpha_{i}(\mathbf{f},\mathbf{d}))\right|$
with composite parameters $\mathbf{f=}f_{1},f_{2}..f_{n}$ and $\mathbf{d=}d_{1},d_{2}..d_{n}$.
 $s_{i}$ and $\alpha_{i}$ represent the normalized solar spectrum data \cite{RenewableDataCen}
and effective  nanocomposite absorption at a given wavelength $i$ respectively. Since the majority
of the solar energy is concentrated in the visible band (see Fig. \ref{fig:optimizedspec}),
we only define this {}``energy'' for $i$ spanning from 350 to 800 nm. This choice of $E$
allows for the absorption profile that most closely resembles the shape of the solar spectrum
to be determined. The algorithm begins by initializing the system to some initial state $E^{s}(\mathbf{f},\mathbf{d})$.
A neighboring state $E^{n}$ is called by using the condition\begin{equation}
E^{n}(\mathbf{f^{n}},\mathbf{d^{n}})=E^{s}(\mathbf{f+\mathbf{\Delta\mathbf{f}}},\mathbf{d+\mathbf{\Delta\mathbf{d}}})\label{eq:neighbor}\end{equation}

where $\mathbf{\Delta\mathbf{f}}=f_{max}\mathbf{x}$ with $\mathbf{x}$ being a vector of dimension
$n=3$ with each component taking a random value between $\left\{ -1,1\right\} $. $f_{max}$
represents the maximum magnitude the volume fraction of a particular species may step. Similarly,
$\mathbf{\Delta\mathbf{d}}=d_{max}\cdot\mathbf{x}$ with $d_{max}$ representing the maximum
magnitude the diameter of the particles corresponding to a particular species may step. Here,
 values of $f_{max}=0.1\%$ and $d_{max}=0.1$ nm provided sufficient finesse in moving through
the search space. Next, the neighboring state is compared with the best state encountered thus
far $E^{b}$ (where $E^{b}$ was originally initialized to the same state as $E^{s}$). If $E^{n}<E^{b}$
then $E^{b}$ is set equal to $E^{n}$. Then, the algorithm must decide if this neighbor state
will become the preferred state for the system. To accomplish this, a Boltzmann-type probability
$P$ analogous to classical statistical physics is calculated-- namely \begin{equation}
P=\exp(\frac{-(E^{n}-E^{s})}{kT})\label{eq:prob}\end{equation}

where $T$ is the annealing parameter analogous to temperature and $k$ is a constant used to
refine the annealing schedule. Note that if $P>1$ then $P$ is simply reassigned to 1. $P$
is then compared to a random number $x$ between $\left\{ 0,1\right\} $. If $x<P$ then $E^{n}$
is accepted as the new system state and $E^{s}(\mathbf{f},\mathbf{d})=E^{n}(\mathbf{f^{n}},\mathbf{d^{n}})$.
Notice from Eq. \ref{eq:prob} that if $E^{n}<E^{s}$, then the neighboring state is always accepted
as the new system state. The process of calling neighbor states and deciding whether or not to
accept them as system state is repeated over $C$ cycles. This represents a random walk of $C$-steps
through the $\left\{ \mathbf{f,d}\right\} $ parameter space. For optimization calculations performed
here, $C$ was chosen to be $500$  as for values greater than this no appreciable difference
in the results was observed. At each iteration of the algorithm, only values in the dilute regime
and outside the realm of quantum effects were accepted (i.e. $\sum f_{n}\leq0.1$ and $d_{n}\geq10$
nm were enforced at each step). In addition, the maximum particle size was constrained to $d_{n}\leq30$
nm as mentioned above  and only physically realizable values of volume fractions were permitted
($f_{n}\geq0$). If the algorithm suggested a neighbor state outside the allowable domain, a
large value was assigned to the energy ($E^{n}\sim10^{8}$). This, by virtue of Eq. \ref{eq:prob},
gives a very low probability for such states to be   accepted. Finally, $T$ is reduced and the
entire process repeated again. $T$ is reduced according to a prescribed schedule until it is
nearly equal to zero after $N$ iterations. Notice that as $T$ is slowly reduced the system
begins to accept lower and lower energy configurations until the it is forced to into a (global)
minimum. For calculations conducted here, the annealing schedule prescribed was $T_{N}=(1-\mu)T_{N-1}$
where $\mu=0.99$ using $N=1000$ iterations. The initial value of $T$ was chosen so that the
probability of the algorithm proceeding from a lower state to a higher state and vice versa was
approximately the same. This ensured that the search space was relatively large during the initial
stages of the anneal. A value of $T=100$ with $k=0.1$ was used for simulations here. Typically,
$\mathbf{f=}<3.0\%,3.0\%,3.0\%>$ and $\mathbf{d=}<15.0nm,15.0nm,15.0nm>$was taken as an initial
guess. When numerical parameters were determined such that different simulation runs yielded
the same answer (within a tolerance of $\Delta E\sim10^{-7}$), different initial states where
chosen and simulations ran again. This was done to verify that the algorithm had, in fact, determined
the system configuration yielding the global minimum. The optimized configuration is presented
in Fig. \ref{fig:optimizedspec} with material parameters 1.80\%Cu:0.35\%Ag:6.4\%Au with respective
particle diameters of 10.1 nm, 29.6 nm and 10.0 nm.

\section{CONCLUSION}

In summary, we have extended our homogenization procedure to quaternary nanocomposites. This
procedure yields an analytic expression that acts as a constraint for tuning the spectral characteristics
of the nanocomposite system. Via manipulation of metal volume fraction and nanoparticle diameter,
one is able to tune the spectral response over the degree of freedom allowed by the nanocomposites'
constituents. We note the selection of Ag, Au, and Cu in SiO$_{\text{2}}$ allows for a controllable
broadband filter with a wide degree of freedom over a spectral range of 350-800nm. Furthermore,
we have shown how a nanocomposite of this type maybe optimized to offer a promising potential
application as an absorption coating on solar devices. 

\begin{acknowledgments}
R.K. and R.S. acknowledge support by the National Science Foundation through CAREER Grant No.
DMI-0449258 and No. CTS-0335348, respectively.
\end{acknowledgments}
%\bibliographystyle{spiebib}
%\bibliography{opticsfilter,SPIEref}

\pagebreak


\end{document}